\documentclass[fleqn,10pt]{wlscirep}
\usepackage[normalem]{ulem}

\newcommand{\rev}[1]{\textcolor{black}{ #1 }}
\begin{document}

\title{
\rev{Modelling the  evolution of transcription factor binding preferences in
complex eukaryotes}
}

\author[1,*]{Antonio Rosanova}
\author[1]{Alberto Colliva}
\author[1]{Matteo Osella}
\author[1]{Michele Caselle}
\affil[1]{Department of Physics and INFN, Universit\`a degli Studi di Torino, via P.Giuria 1, I-10125 Turin, Italy}

\affil[*]{antonio.rosanova@unito.it}



\begin{abstract}
Transcription factors (TFs) exert their regulatory action by binding to DNA with specific sequence preferences. 
However, different TFs can partially share their binding sequences due to their common evolutionary origin. 
This ``redundancy'' of binding defines a way of organizing TFs in ``motif families'' by grouping TFs with similar binding preferences. 
Since these ultimately define the TF target genes, the motif family organization entails information about the structure
of transcriptional regulation as it has been shaped by evolution. 
Focusing on the human TF repertoire, we show that a one-parameter evolutionary model of the Birth-Death-Innovation type can explain the TF empirical repartition 
in motif families, and allows to highlight the relevant evolutionary forces at the origin of this organization. 
Moreover, the model allows to pinpoint few deviations from the neutral scenario it assumes: three over-expanded families (including HOX and FOX genes), 
a set of ``singleton'' TFs for which duplication seems to be selected against, and a higher-than-average rate of
diversification of the binding preferences of TFs with a Zinc Finger DNA binding domain. Finally, a comparison of the TF motif
family organization in different eukaryotic species suggests an increase of redundancy of binding with organism complexity.
\end{abstract}

\flushbottom
\maketitle
%
%
\thispagestyle{empty}
\section*{Introduction}
Transcriptional regulation plays a crucial role in most physiological processes, ranging from cell homeostasis 
to differentiation~\cite{accili2004foxos,bain1994e2a,dynlacht1997regulation}, and its disregulation is often implicated in pathological processes such as cancer~\cite{Furney2006}. 
Mainly thanks to transcriptional regulation, species with highly similar genome sequences can have radically different 
expression patterns and as a consequence very different phenotypes~\cite{bustamante2005natural,de2008patterns,lopez2008functional, Voordeckers2015}.  
 Therefore, deciphering the mechanisms of evolution of transcriptional regulation is  a core 
part of modern evolutionary biology~\cite{Teichmann2004,MadanBabu2006, Cordero2006, Enemark2007, Pinney2007, Aldana2007, Crombach2008, Nowick2010}.
\newline
\noindent
Transcriptional regulation is mainly controlled by a class of proteins known as transcription factors (TFs) 
which are characterized by the presence of at least one DNA binding domain (DBD), 
i.e., a structural domain able to mediate the TF-DNA interaction. Through this protein-DNA interaction, 
TFs can recognize their target genes and induce or repress their transcription.          
The set of TFs with their corresponding targets ultimately define the complex network of regulations  
that orchestrates the organism gene expression program.    
Therefore, evolutionary changes in the TF repertoire and/or in their sequence binding preferences 
can  induce large-scale alterations in the gene expression program, thus representing a primary potential source of phenotypic variation and evolution.  
\newline
\noindent
Gene duplication and gene loss are main  drivers of genome evolution and thus also of the TF repertoire~\cite{Ohno1970,Zhang2003,Demuth2009}. 
For example, in eukaryotes around the 90\% of genes is the result of an event of gene duplication~\cite{Voordeckers2015,Conant2008,Teichmann2004,Lynch2000}. 
Moreover, changes in gene copy numbers  play a role in evolutionary adaptation comparable to the role of sequence alteration through mutations~\cite{Demuth2009}, and  
this may be particular true for the evolution along the human lineage~\cite{Demuth2009}, which will be the main focus of this paper. 
Indeed,  gene gain and loss seem to account for a large part of the human/chimpanzee genetic divergence~\cite{Britten2002,Cheng2005}. 
These  basic evolutionary moves of  duplication and deletion can significantly alter the transcriptional regulatory network  by expanding or reducing the number of TFs 
with certain specific binding preferences. After duplication of a TF gene, the two resulting gene copies are likely redundant.  
In fact, initially the two TFs share the same sequence, including the DBD sequence that encodes their binding preferences, and thus they also bind to the same target genes. 
Subsequently, mutations in the DBD sequence can eventually induce one of the TF copies 
to switch to regulating different target genes~\cite{Perez2014}, thus resolving the initial redundancy.  
Alternatively, the regulatory redundancy may be retained  to increase the network robustness~\cite{Gu2003},  
or  the combinatorial  complexity of regulation  if the two TFs continue to regulate the same set of target genes 
but evolve to respond to different cellular signals or to interact with different proteins~\cite{Baker2013,Zhang2003}. 
The organization of TFs in ``families'' collecting TFs with the same binding preferences, thus putatively TFs with highly overlapping sets of target genes, 
should carry signatures of the  evolutionary forces in action. For example, a duplication event  expands a TF family, while the progressive sequence divergence of a 
TF may give rise to a new TF family able to  recognize a significantly different set of target genes. These dynamics  
could  be typically dominated by neutral  evolution, but the TF organization may also conceal hallmarks of adaptive selection  that, for example,  drove
the over-expansion of specific TFs or their functional diversification. 
\newline
\noindent
The goal of this paper is precisely to design a method to address quantitatively the evolutionary dynamics  that shaped the TF repertoire 
and their TF binding preferences. In order to do so, we first propose a method to organize TFs in families based on their binding preferences that we call ``motif families''.   
Second, we introduce a simple stochastic model of neutral evolution  based on the duplication-and-divergence dynamics described above that can be treated analytically and with stochastic simulations. 
The model introduces a neutral scenario for the distribution of sizes of the TF families able to explain the general empirical repartition of TFs in motif families in human. 
At the same time, a quantitative theoretical framework allows to pinpoint specific  deviations from the neutral expectations that can be the result of selection. 
The model also introduces a natural measure of  TF binding redundancy, and by comparing several eukaryotic model species a striking evolutionary trend can be identified.   
\section*{Results}
\subsection*{Organization of TFs in motif families}
Although the number of TFs may vary substantially  from genome to genome, the number of distinct DBD types is small.
\rev{In fact, a previous study~\cite{weirauch2014determination} distinguishes} just barely one hundred sequence-specific DNA-binding domains.
The metazoa-specific set of DBDs is limited to a few dozens. 
Such a classification is perfectly suited to identify long-term patterns of duplication and conservation, 
but it is too coarse-grained  to capture the fine changes in regulation which occur on a much faster evolutionary time scale. 
Indeed, just a few single-nucleotide mutations in the DBD active site are enough to \rev{modify} the binding preferences, without a significant change of the DBD structure.
To highlight these fine changes of binding preferences a ``PWM based'' classification of TFs is mandatory. 
Such a classification was out of reach up to a few years ago, due to the uncertainty in PWM definition (above all for paralogous TFs!), but can be now addressed in a reliable way 
thanks to the recent experimental and computational progress in PWM reconstruction~\cite{weirauch2014determination}.
Leveraging on this remarkable progress, we propose here a classification of TFs based on their binding preferences, following the approach of Jolma et al.~\cite{jolma2013dna}.
The result of this classification is an organization of TFs in what we call \textit{motif families}, which group together TFs associated to the same PWM (\rev{see below for a more precise definition}).
This organization in motif families is a sub-partition of the DBD classification, 
which is expected to be more closely related to the TF regulatory potential and thus to evolutionary forces which shaped the regulatory network.
\rev{This paper proposes a model of the evolutionary process at the origin of this TF organization, which is essentially the following}.   
After a duplication event, TFs in the same DBD class are in the same motif family. 
Mutations may drive a TF out of its motif family, giving rise to a new motif family, but remaining in the same DBD one.
\begin{figure}[!ht]
\begin{center}
\includegraphics{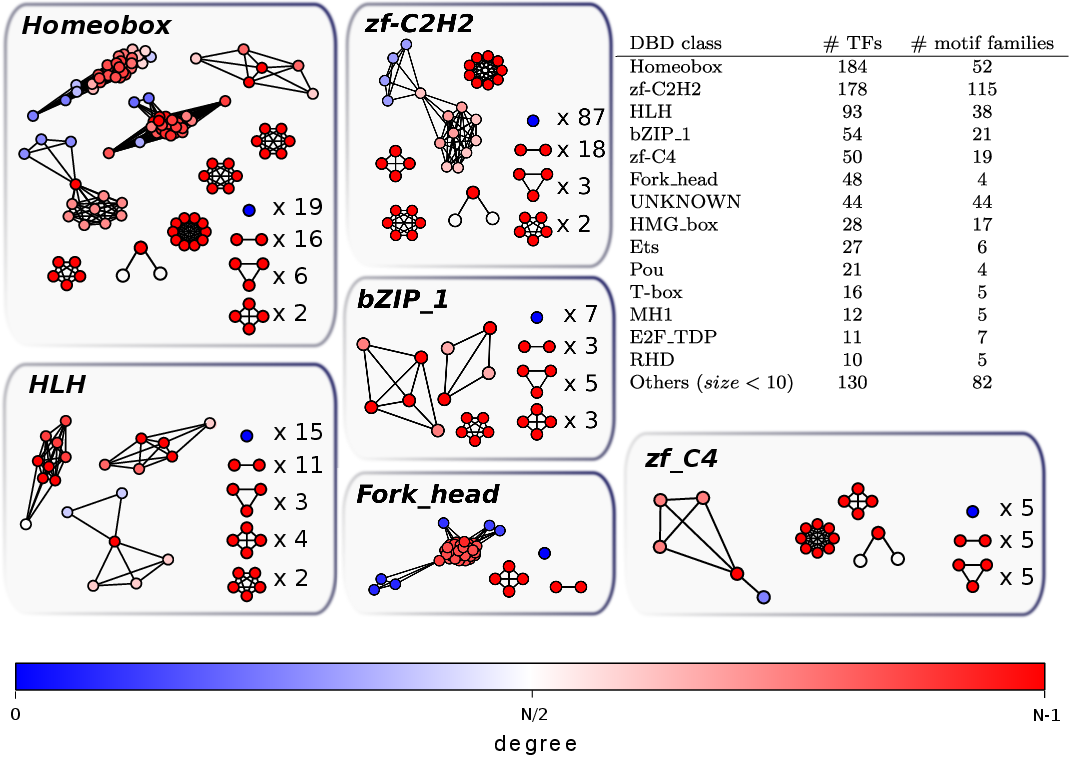}
\end{center}
\caption{{\footnotesize
{\bf Graphical representation of motif families.} {The table summarizes the organization of the DBD families in motif families.} 
Each vertex is a transcription factor, while links connect two TFs if they share at least one Position Weight Matrix ID. 
Colors identify the node degree with a color code (reported in  the legend) spanning from blue, corresponding to degree 0 (isolated nodes), to red, 
corresponding to the maximal  degree, i.e.,  the vertex is connected to all other nodes in the family. The circular layout highlights those families that are cliques.
See Supplementary Material for reference to detailed family composition.}}
\label{fig:1}
\end{figure}
\newline
\noindent

\rev{We based our analysis on the PWM classification proposed in a previous work~\cite{weirauch2014determination}.
In this classification, each TF is associated to a set of PWMs obtained with different experimental techniques or inferred on the basis of DBD homology~\cite{weirauch2014determination}. 
This homology-based inference allows to associate a PWM experimentally found for a specific TF to other TFs in the same DBD class that show a particularly 
high homology in the DNA binding domain~\cite{weirauch2014determination}. 
In principle, one could combine these different PWMs to construct a single comprehensive PWM for each TF, but the  
different methods used to obtain them (with different resolution power) suggest to avoid this merging procedure. 
Instead, the PWM/TF association can be represented as a biparite network with two classes of nodes (TFs and PWMs) and links between TFs and PWMs if they are associated in the CIS-BP database. 
By construction, in this network there are no direct links between PWMs. 
It is easy at this point to construct the ``TF projection'' of this bipartite network, which is composed only by the TF nodes with links connecting two TFs
if they are associated to least one common PWM.
The network defined in this way is characterized by several disconnected
components of high link density, each of which defines a motif family (Figure~\ref{fig:1}).
%
Most of these components are cliques, i.e., groups of TFs with at least one PWM in common among all the members.
Figure~\ref{fig:1} shows that most of the DBDs families are split in smaller more specific motif families.
The ``splitting rule'' turns out not to be uniform, as some DBD classes appear more inclined to diverge than others.
Three examples of the splitting of DBD families in motif families are discussed in detail in Section 5 of the Supplementary Material. 
%
Figure~\ref{fig:2}  reports the size distribution of motif families. It is worth noting the large number of motif families of size 1, representing isolated TFs.
The size distribution in Figure~\ref{fig:2} is the observable that we aim to explain in terms of a simple evolutionary model.
}

\rev{Due to the organization of the CIS-BP database, the TF-TF links that we find with our procedure are mainly due to the ``inferred'' TF-PWM associations of the CIS-BP database,
and thus are related to the level of homolgy between the DBDs of the two TFs. The main assumption of the CIS-BP inference procedure (and thus of the motif family definition) is that 
high levels of DBD homology should imply high similarity of the corresponding PWMs. In order to assess the robustness of our construction with respect to this assumption, 
we tested how much the proposed motif families organization would be affected by the inclusion of additional links between TFs on the basis of a direct measure of similarity between their PWMs.
The procedure for this robustness test is explained in detail in Section 6 of the Supplementary Material.  
The Jaccard index can be used as a measure of similarity between each pair of PWMs and thus indirectly between the binding preferences of the corresponding TFs. 
The TF-TF network defined above 
can thus be expanded by progressively adding links as the critical threshold for this similarity index is lowered. 
It turns out that  most of the new links coincide with already existing ones or simply join TFs already belonging to the same motif family. 
Only when the  thresholds of similarity between PWM approaches really low values, links connecting TFs belonging to different families start to appear. 
This result show the close link between DBD homology and PWM similarity, and supports the robustness of the motif family organization used here. 
}
\begin{figure}[!ht]
\begin{center}
\includegraphics[width=0.5\textwidth]{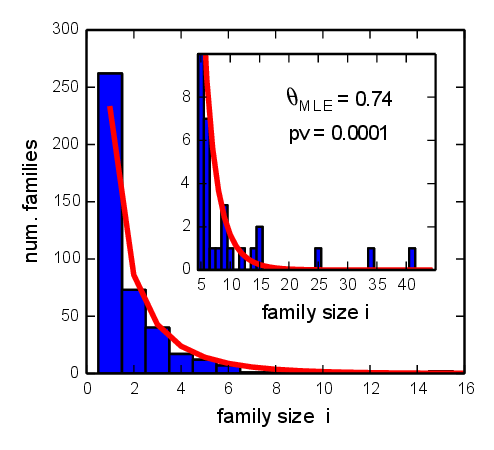}
\end{center}
\caption{{\footnotesize
{\bf Size distribution of motif families in human.}   The distribution accounts  for 906 human TFs,  organized in 424 families whose members share at least one PWM with at least
another member.
The inset is a zoom on the range of sizes $>5$. The red-line is the best-fit model according to maximum likelihood estimation, which has a 
goodness-of-fit p-value $p< 0.0001$. The model captures the general trend, but clearly underestimates the number of families of size 1 and does not predict the presence 
of the largest families.  }}
\label{fig:2}
\end{figure}
\subsection*{The Birth-Death-\textit{cis-}Innovation model}
The model we propose belongs to the general class of Birth-Death-Innovation models (for a thorough introduction see~\cite{karev2002birth}).
The focus  of these models is on systems in which individual elements are grouped into families whose evolution is ruled by the dynamics of their individual members.
These models typically include the elementary processes of family growth via element duplication (gene duplication), element deletion as a result of inactivation or loss
(negative gene mutation), and innovation or emergence of a new family (neutral/positive gene mutation). 
All these processes are assumed to be of Markov type and the corresponding rates are assumed to be constant in time.
\newline
\noindent
It can be argued that the total number of TFs has been tuned to an optimal one to address in the most efficient way the regulatory needs of the organism. 
In fact,  it has been observed  that an upper bound must exist on the total number of TFs  to ensure an
optimal coding strategy in which misrecognition errors are minimized~\cite{itzkovitz2006coding}.
Since we aim to describe only the evolution of the TF regulatory strategies in complex eukaryotes,  
we shall assume that the mean number of TFs is essentially constant over time and stably close to the optimal value.  
In fact, the dynamics in which we are interested in is the evolution of the binding preferences of these TFs, 
which is presumably acting on a faster timescale with respect to the changes in the TF total number. 
This assumption of a separation of time scales is in line with the notion of punctuated equilibrium often implied in 
several evolutionary models\cite{koonin2002structure}:
long period of stasis are punctuated by short bursts of evolutionary activity that involve radical alterations of the duplication and elimination rates. 
Between these periods of drastic changes, the system seems to rapidly relax to equilibrium. 
The assumption of equilibrium justifies the assumptions of rates constant in time and an approximate balance between the mechanisms generating an inflow and an outflow of genes,
so that the total number of TFs stays constant by mean.
\newline
\noindent
We introduce the dynamic of cis-innovation that makes a TF become the seed of a new family.
Given that the repertoire of DBDs in higher eukaryotes is remarkably conserved over the last 600 million years, cis-innovation stands 
as the driving force of TF innovation on the time scale of PWM evolution we are interested in.
In fact,  our model description focuses only on the ``late'' stage of TF evolution in metazoans,  in which very few new DBDs, and thus new motif families, are created \textit{de novo}.
\newline
\noindent
In conclusion, we shall evaluate the family size distribution as the stationary state of a process of duplication, deletion and divergence, where the total 
number of TFs is essentially stable.
To introduce the  model in more detail, let us define as ``class $i$'' the set of all families of size $i$. 
Let $f_i$ be the number of families in the $i$-th class,  $M$ be the total number of classes $i=1....M$ (or the maximum size of a family), 
and $N$ the total number of elements, thus representing also the extreme value for $M$.  
Acting at  the ``local'' level on individual elements, the evolutionary dynamics shapes ``globally'' the system relocating a family from class $i$ to 
class $i+1$ in case of  duplication (or to class $i-1$ in case of removal).
Typically,  BDI models\cite{novozhilov2006biological,fenner2005stochastic,lagomarsino2009universal} 
introduce innovation in the model only as a constant inflow in the class 1 
due to \textit{de novo} emergence of a new family (increase of $f_{1}$ by 1).
As discussed above, we propose a generalization of the model by introducing also cis-innovation,  
in which an element of a family in class $i$ mutates and gives rise to a new family. 
This results in the relocation of that element in class 1 and of its original family in class $i-1$ (i.e. a decrease of $f_i$ and increase of $f_{i-1}$ and $f_{1}$ by 1).
Let $\lambda$, $\delta$, $\nu$ and $\mu$ be the rates of element birth, death, \textit{de novo}-innovation and \textit{cis}-innovation respectively. 
Solving the master equations at the steady state (see the Materials and Methods section) one finds:

\begin{align}
 f_i=\frac{\nu+\mu N}{\lambda}  \frac{\theta^{i}}{i},
\label{eqx}
\end{align}

where $\theta=\frac{\lambda}{\delta+\mu}$. \\
The corresponding probability distribution $p_i$ can be found straightforwardly by normalization: 

\begin{align}
 p_i=\frac{f_i}{\sum_{i}{f_i}}= \frac{1}{\sum_{i}{\frac{\theta^{i}}{i}}} \frac{\theta^{i}}{i}.
\label{eqx2}
\end{align}

\noindent
A few comments are in order at this point:
\begin{itemize} 

\item
The normalized solution in Equation~\ref{eqx2} gives a one-parameter prediction of the size distribution of motif families. 
The functional dependence on $\theta$  is equivalent to the one that can be obtained with standard BDI models\cite{karev2002birth}, i.e., with  de novo innovation as the only source of innovation.
However, our generalized model suggests a different interpretation of the parameter.
In fact, $\theta=\frac{\lambda}{\delta+\mu}$ and thus its value depends on the  rate of cis-innovation. 
\item
The steady state condition is $\frac{df_i}{dt}=0$ $ \forall i$,  implies that  the total number of elements $N=\sum_{i}^{M}{if_i}$ is constant over time. 
This condition translates into the parameter constraint $N(\delta-\lambda)=\nu$. 

\item
As previously discussed,  we expect $\nu$ to be very small in our case (i.e., negligible de novo innovation), and accordingly we shall approximate $\nu\to 0$ in the following. 
We shall further verify ``a posteriori'' the validity of this approximation using an independent analysis on the evolution of TFs in different lineages (see below).
In this regime, the stationary condition simplifies to a balance between duplication and deletion rates $\lambda =\delta$,  
and  $\theta\simeq \frac{1}{1+\mu/\lambda}$. 
Therefore, the deviation of $\theta$ from 1 allows to directly estimate the magnitude of  $\mu$ with respect to $\lambda$, i.e., 
the relevance of cis-innovation with respect to the birth/death rate. 
As we will see below, a comparison with the data in the human case  supports a value of $\theta \sim 0.73$, thus highlighting the important role that cis-innovation had in the recent
evolution of the eukaryotic TF repertoire. 
Moreover, within this approximation, also the family distribution in Equation~\ref{eqx}  can be written in a very simple and compact form:
 
\begin{flalign}
  f_i=N \frac{\mu}{\lambda} \left(\frac{\lambda}{\delta+\mu}\right)^i\frac{1}{i}=N (1-\theta)  \frac{\theta^{i-1}}{i}.  
\label{eq1}
\end{flalign}

\item
An analytical estimate of the number of classes $F=\sum_{i}^{M}{f_i}$  in which the $N$ elements are organized when the
dynamics reaches equilibrium can also be calculated as:

\begin{flalign}
  \frac{F}{N}=\frac{1-\theta}{\theta}\sum_{i}^{M}{\frac{\theta^i}{i}} \simeq  \frac{\theta-1}{ \theta} ln(1-\theta)
  \label{eq3}
\end{flalign}

This represents the neutral model prediction on the number of motif families given a set of $N$ TFs subjected to the described BDI dynamics.  
\end{itemize}

\subsection*{The model can explain the core of the size distribution of motif families and identifies two main deviations}
\begin{figure}[!hb]
\begin{center}
\includegraphics[width=0.8\textwidth]{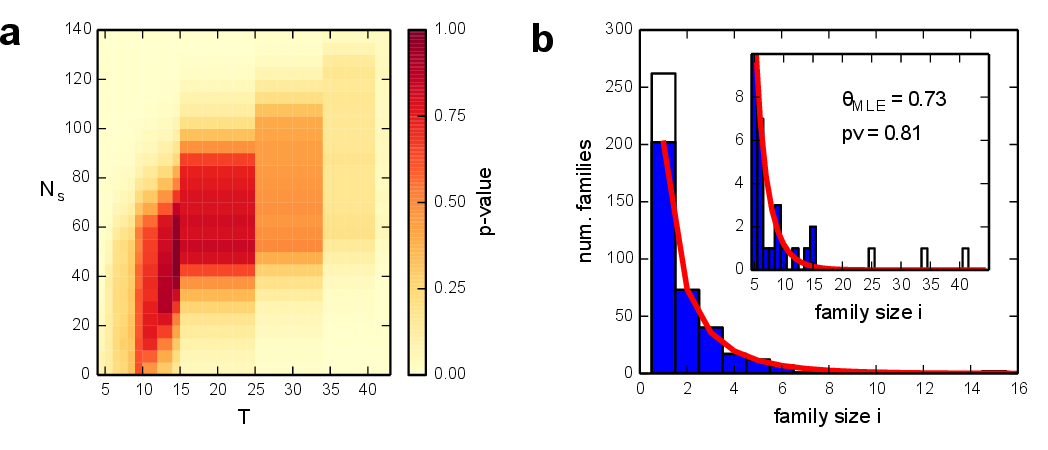}
\end{center}
\caption{{\footnotesize
{\bf a) Heatmap for the goodness-of-fit p-value  as the data sample is reduced}. On the x-axis, $T$ indicates the threshold in size above which families are excluded from the sample. 
On y-axis $N_s$ indicates the number of families of size one excluded from the sample. An increase in $T$ or $N_s$ reduces the sample size in analysis by reducing the number of TFs considered. 
For each sample size a goodness-of-fit test for the best-fit model was performed and the corresponding p-value is reported with the color code in the legend.   
Considering a p-value of $0.75$ as the acceptance limit  identifies  $T=25$ as the size threshold at which the fit is acceptable. This  corresponds to the exclusion of  the
three largest families. For such a threshold $T$, the optimal values for the p-value are reached for values of $N_s$ in the range $40<N_s<80$.
{\bf b) Size distribution of motif families for the reduced sample}. The filled distribution represents the motif family size distribution for the reduced dataset, while the original empirical 
distribution of Figure 2 is reported with the unfilled bars. The inset shows a zoom on the range of sizes $>5$. The line represents the prediction of our model with the best fit choice of the parameter 
$\theta$ , which turns out to fit very well the data contained in the reduced sample with a goodness of fit p-value $p\sim  0.8$.
The best fit value $\theta=0.73$ does not differ substantially from the value that is obtained by fitting the whole empirical sample as in Figure~\ref{fig:2}. 
  }}
\label{fig:3}
\end{figure}

The distribution predicted by our neutral evolutionary model (Equation~\ref{eqx2}) can be compared with the empirical TF organization in motif families.
The procedure to extract this empirical distribution is explained in the Materials and Methods section in detail. 
This comparison can be quantified by estimating the best fit value of the parameter $\theta$ with a Maximum Likelihood method and 
a p-value associated to the quality of the fit using
a \textit{goodness-of-fit} test  based on the Kolmogorov-Smirnov statistics (Materials and Methods). 
Although the central part of the size distribution seems well captured by the theoretical model, a direct fit of the whole distribution 
gives very low p-values ($\text{p-value} < 10^{-3}$, see Figure~\ref{fig:2}). This poor p-value shows the presence of significative deviations with respect to our random null-model. 
These deviations can be easily identified looking at Figure~\ref{fig:2}.  
They are located at the two ends of the distribution and involve a few of the largest families and the smallest ones (i.e., families of size 1). 
Using the KS test and a p-value threshold for acceptance of 0.75, we can identify in a quantitatively and consistent way the fraction (about 25\%) of 
isolated TFs and the number (three) of the largest families which account for most of the deviations from the null model (Materials and Methods and Figure~\ref{fig:3}).
\newline
\noindent
If we subtract from the whole distribution these two tails (for a total of $\sim 150$ TFs, i.e. about 16\% of the total number of TFs in analysis), we eventually find  a 
remarkable agreement between  the model predictions and experimental data ($\text{p-value} \sim 0.8$, see Figure~\ref{fig:3}).
Therefore, the ``core'' of the distribution is well described by the exponential-like solution of Eq. (\ref{eq1}), while deviations are due to few families that can be isolated and studied in detail.    
This suggests that the evolution of a large portion of the TF repertoire in higher eukaryotes was driven by a neutral stochastic process of the BDI type with only two exceptions: 
an excess of isolated TFs and three large families which on the contrary are characterized by a strong level of duplication without innovation.       
Let us address in more detail these two deviations.

\subsubsection*{Single copy transcription factors}
The fitting procedure allows us to obtain a rough estimate of the fraction $N_s$ of size 1 families which are not explained by our theoretical description.  
This number is in the range $40 <N_s< 80$ , i.e, in between 20\% and 30\%  of the total number of size 1 families (Materials and Methods and Figure~\ref{fig:3}).
The emergence of a size 1 family in our model description can come from de novo innovation  or  from duplication of an existing TF, 
followed by a cis-innovation event that defines a new PWM. 
We argued that de novo innovation is negligible in our case of study, so we expect that most of the isolated TFs are the result of a previous duplication event. 
In this scenario, they should share their DBD at least with the TF they duplicated from, and  we verified that indeed empirically this is the case for the  majority of isolated TFs, 
thus supporting our model description. 
However, some isolated TFs  have a DBD which is not shared with any other TF (12 in our sample)
 or are characterized by a DBD which is classified as 'UNKNOWN' (44 in our sample), so also potentially unique. 
The presence of these isolated TFs with unique DBDs can be explained by the two following mechanisms. 
\begin{description}
\item{{\sl Newly acquired DBDs.}}
A few of them are due to actual recent de novo innovation events, thus introducing new DBDs in the last period of post-metazoan evolution. These  ``recent'' TFs 
appear in our analysis most likely as size 1 families only because they had not time to enter into the duplication process. 
Looking at the orthology maps  we can rather easily identify these DBDs and the corresponding TFs (see Supplementary Material and below) 
which turn out to be very few, thus supporting ``a posteriori'' our $\nu=0$ approximation.

\item{{\sl Singleton genes.}}
The majority of excess isolated TFs are most probably \textit{singleton genes} for which duplication is peculiarly avoided. 
The existence of this class of genes has been recently proposed~(\cite{carroll2008evo,d2011modification}).
They are supposed to be ancestral genes of prokaryotic origin, addressing basilar functions and requiring a fine-tuning of their abundances,  
thus making their duplication particularly detrimental. 
They would be the result of a selective pressure to avoid duplication, and thus, by definition, cannot be explained by our neutral model. 
\end{description} 

\noindent

Since singleton genes are not included in our  model, they are good candidates to explain the excess of isolated TFs in Figure~\ref{fig:2}. 
To distinguish between putative singleton genes and recent genes in  the motif families of size 1, 
we analyzed their  evolutionary origin.  
More specifically, we manually inspected the taxonomic profiles of these 56 TFs in the EggNOG database~(\cite{huerta2015eggnog}): 
16 of them have a putative origin at the Last Universal Common Ancestor (LUCA), i.e. they are shared among
bacteria, archaea and eukarya;  25 are in common among all eukarya, 4 among opisthokonta, 3 among metazoa and 8 have a post-metazoan origin.
Therefore,  at least 41 of these TFs  have a very ancient origin (LUCA + eukarya) and could well be examples of  ``singleton'' TFs,  
while 8 are instead of very recent origin (post-metazoan, but 4 of them are shared only among euteleostomi) and are thus likely to be ``recent'' TFs.
These recent TFs constitute less than the $1\%$ of our sample, supporting ``a posteriori'' the $\nu=0$ approximation.

To find additional evidence that these  41 ancient TFs can be bona fide ``singleton genes'', we queried the  NGC5.0 database~(\cite{An2016}), 
which provides information about the gene duplicability for a large set of cancer genes. 
14 of our putative singletons are present in this collection, and 12 of them show indeed no evidence of duplicability (at 60\% coverage), thus supporting their ``singleton'' nature.
It is interesting to notice that the overall number of putative singletons (41 genes) 
is compatible to the size of the deviation from the random null model (40 $<N_s< $80)  observed in our best fit tests.
\rev{An example of a DBD family (the IRF family) giving rise to a set of motif families of size 1 is discussed in detail in the Supplementary Material (Section 5).}

\subsubsection*{Over-expanded families}
Our analysis singles out also three over-expanded families. The over-expansion can be due to two parallel mechanisms: 
an enhanced rate of duplication and/or a decreased rate of cis-innovation. 
Looking at the three over-expanded families, three very homogeneous groups of TFs can be recognized: 
the FOX family (size 41), the HOX family (size 34) and another homeobox family (size 25).
These three families are good examples of the two mechanisms mentioned above. 
The HOX family contains TFs well known for their role in morphogenesis and animal body development~(\cite{pavlopoulos2007hox}).
Also TFs in the other over-expanded homeobox family show enrichments for GO annotations related to \textit{morphogenesis}, 
\textit{development} and \textit{pattern specification}, as reported in Table~\ref{tab:1}.
These two families may well represent  cases of  positive selection for duplication and subsequent fixation. 
Due to their crucial role in morphogenesis, these TFs could have been retained in multiple redundant copies to ensure proper response under radically changing conditions.
\newline
\noindent
The third family, which is the largest one, collects most of the FOX  (Forkhead box) TFs present in the sample. 
TFs belonging to this family are known to be ``bispecific'', i.e. 
they recognize two distinct DNA sequences~(\cite{nakagawa2013dna}), and for this reason they play an important and peculiar role in the regulatory network of metazoans~(\cite{nakagawa2013dna}). 
While their over-expansion can be due to positive selection for functional reasons, 
their unique feature of bispecific binding could suggest that innovation 
is particularly  difficult for these TFs. In fact,  bispecifity is likely to impose stronger constraints, from a structural point of view, 
than those imposed on other TFs. In this perspective, it is interesting to stress the different distribution of Forkhead and Homeobox genes in motif families.  
Almost all the Forkhead genes are collected in this single large motif family, suggesting no cis-innovation events that would have moved  some of these genes in families of other sizes. 
Only 6 Forkhead TFs are present in other motif families. 
On the other hand,  Homeobox genes, besides the two main families discussed above, are dispersed in several other motif families, thus are associated to a variety of PWMs.  
This difference suggests that  duplication of Homeobox genes has been positively selected at a certain time point  probably because of  
their crucial role in the development of multicellular organisms (see Table~\ref{tab:1}), but cis-innovation have progressively changed their binding preferences. 
On the other hand, very few events of cis-innovation are associated to FOX genes that indeed ``accumulated'' in a single motif family. 
These interpretations of the possible evolutionary origins of the over-expanded motif families will be addressed in more detail in the next section.
\begin{center}
\begin{table*}[ht]
{\footnotesize 
\hfill{}
\begin{tabular}{l|cr}
GO biological process complete     &     Fold Enrichment &    p-value \\
\hline
embryonic skeletal system development (GO:0048706)    & 6.92    & 7.87E-18 \\ 
skeletal system development (GO:0001501)   & 4.21    & 8.44E-15 \\ 
embryonic skeletal system morphogenesis (GO:0048704)    & 7.18    & 1.20E-14 \\ 
skeletal system morphogenesis (GO:0048705)      &    5.77    & 4.29E-14 \\ 
anterior/posterior pattern specification (GO:0009952)    &    4.91 &    2.55E-13 \\ 
pattern specification process (GO:0007389)    & 3.10    & 3.15E-10 \\ 
regionalization (GO:0003002)            & 3.35    & 7.25E-10 \\ 
embryonic organ morphogenesis (GO:0048562)    & 3.47    & 4.71E-08 \\ 
embryonic morphogenesis (GO:0048598)        & 2.66    & 1.83E-06 \\ 
organ morphogenesis (GO:0009887)        & 2.25    & 4.76E-06 \\ 
chordate embryonic development (GO:0043009)    & 2.59    & 7.91E-06 \\ 
embryo development ending in birth or egg hatching (GO:0009792)    & 2.58    & 9.09E-06 \\ 
embryo development (GO:0009790)            & 2.10    & 2.08E-05 \\ 
embryonic organ development (GO:0048568)    & 2.65    & 4.30E-05 \\ 
anatomical structure morphogenesis (GO:0009653)    & 1.76    & 5.73E-05 \\ 
system development (GO:0048731)            & 1.43    & 7.57E-04 \\ 
anatomical structure development (GO:0048856)    & 1.33    & 8.57E-04 \\ 
multicellular organism development (GO:0007275)    & 1.36    & 1.19E-03 \\ 
animal organ development (GO:0048513)        & 1.49    & 1.97E-03 \\ 
developmental process (GO:0032502)        & 1.30    & 2.77E-03 \\ 
single-multicellular organism process (GO:0044707)    & 1.32    & 4.44E-03\\
\end{tabular}}
\hfill{}
\caption{\footnotesize{{\bf Gene Ontology analysis} of the genes belonging to the two homeobox families of size 25 and 34. 
Only results with p-value below $0.01$ are shown.
The enriched annotations are mainly associated to development and morphogenesis of multicellular organisms.}\label{tab:1}}
\end{table*}
\end{center}

\subsection*{Phenomenology of the splitting of DBD families in motif families}
\begin{figure}[!ht]
\begin{center}
\includegraphics{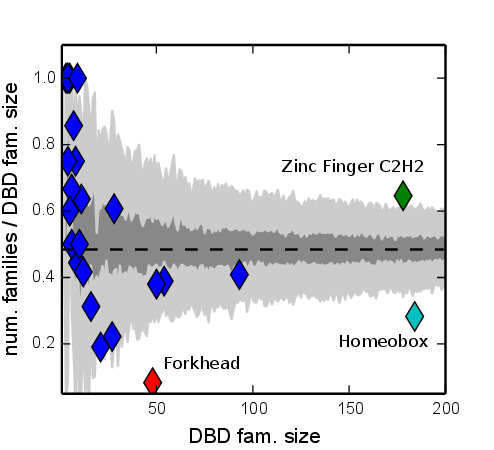}
\end{center}
\caption{{\footnotesize
{\bf Splitting of DBD families in motif families.}  The  ratio $F/N$ is plotted as a function of the number of TFs of the DBD family. 
Each point represents the empirical value for a DBD family, while the dashed line represents the expected value for $\theta=0.74$ as given by Eq. \ref{eq3}.
In order to evaluate the fluctuations on the expectation, we simulated the evolution of $5*10^4$ DBD families,
with starting size ranging from $1$ to $500$, $\theta=0.74$ and $\lambda=\delta$. The two shaded areas correspond to 1 standard deviation and 3 standard deviations from the average.  
 Green diamond: Zinc Finger C2H2 family. Cyan diamond: Homeobox family. Red diamond: Forkhead family.}
}
\label{fig:4}
\end{figure}

So far, we considered the ``global'' distribution of all TFs in motif families. 
However it is also interesting to study separately the behaviour of the different DBD families. 
Each of them can be considered as an independent instance of the evolutionary model described above and it is interesting to 
see if there are significant deviations for specific DBD families with respect to the null model predictions.  
Using as input the value $\theta=0.74$ obtained by fitting the whole set of TFs we obtain from Equation~(\ref{eq3}) a parameter-free prediction for the ratio $F/N$.
 To evaluate also the possible variability of this neutral expectation, we ran $5*10^4$  simulations of the model for different system sizes, 
 corresponding to the different numbers of TFs in the DBD families. We report 
in figure~\ref{fig:4} the comparison of the model prediction (dashed line) and the results of model simulations (shaded areas represent 1 and 3 standard deviations from the average simulated behaviour) with empirical data (symbols).   
While most of the  DBD families do not deviate significantly from the model prediction, three clear  ``outliers'' can be observed. 
The Forkhead and Homeobox DBD families show a smaller than expected number of motif families while the Zinc Finger class of TFs is splitted in more families than expected. These deviations can be traced back to the peculiar features of these DBDs.
In the case of the Forkhead TFs the low value of $F/N$  is 
likely a consequence of the structural constraints typical of the Forkhead DBD which limit the evolvability of the binding preferences leading to
a lower-than-average rate of cis-innovation and thus a smaller number of motif families.
For the Homeobox DBD instead there seems to be  no structural reason for this "freezing" of motif diversity. 
It is tempting to speculate that the low value of $F/N$ is in this case related to the special role played by these genes in the regulatory network. Indeed Hox genes are known to be crucial players of the development of multicellular organisms and it is nice to see how this special role is highlighted by our simple model. 
The other significant deviation from the model prediction concerns a Zinc Finger class of TFs, that appears to have greatly  diversified the TF PWMs. 
The corresponding motif families are not over-expanded, in fact they did not emerge as deviations in the previous analysis (Figure~\ref{fig:2}). 
In fact, the histogram of their motif family sizes (the analogous of Figure~\ref{fig:2} but restricted to Zinc Finger TFs, see Figure~S1) follows reasonably well our null model. 
However, the fitted parameter $\theta=0.56$ is well below the value obtained for all TFs ($\theta=0.74$), thus  confirming again that the rate of cis-innovation for this 
DBD family is higher than the average rate for all TFs. 
Zinc Finger TFs are known to be characterized by multiple tandem C2H2 zinc finger domains. Such modularity enabled a rapid functional divergence among recently duplicated paralogs, as 
each domain in the protein can mutate independently~(\cite{Emerson2009}).  This structural feature is well represented by our simple model. 
\subsection*{TF redundancy of binding increases with organism complexity}
\begin{figure}[!ht]
\begin{center}
\includegraphics{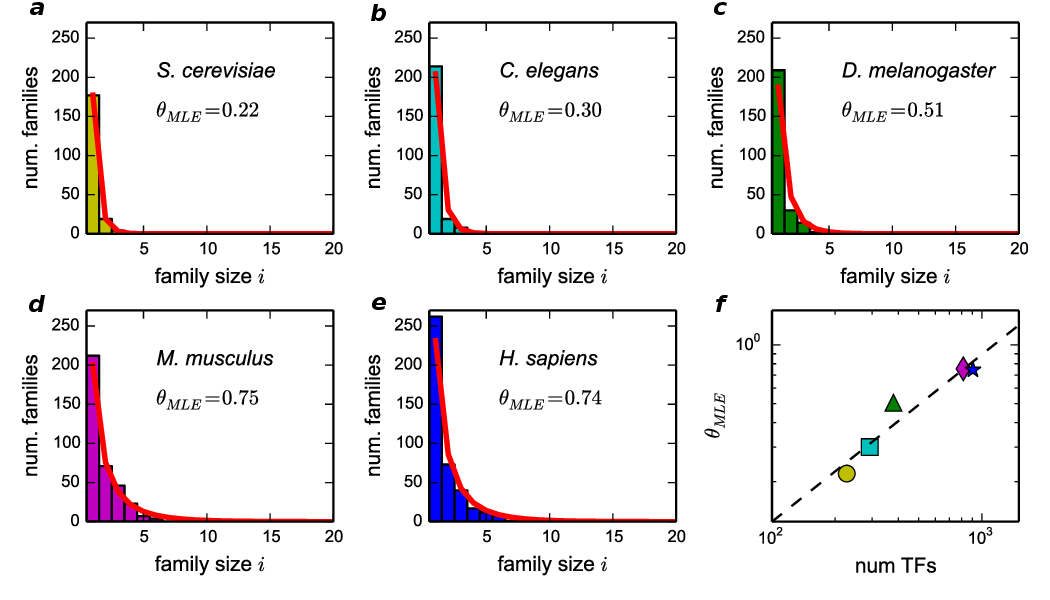}
\end{center}
\caption{{\footnotesize
{\bf Size distribution of TF motif families for different eukaryotic organisms.} {\bf a-e)} We report the distributions and the best fit values of $\theta$ for 
five different organisms of increasing complexity. As for the human case,  the data are taken from the CIS-BP database.
{\bf f)} The bottom-right panel shows how $\theta$ scales with the number of TFs. The red-line is the fit $\theta \sim (\text{\# TFs})^{0.85}$.
}}
\label{fig:5}
\end{figure}

This section addresses the differences in the motif family organization in different eukaryotic species. 
In particular, we focused on model species, which are expected to have  well annotated TF repertoires. 
The same type of analysis presented in Figure~\ref{fig:2}  was performed on the set of TFs of yeast and of three other species of increasing complexity in the animal lineage: 
\textit{C. elegans}, \textit{D. melanogaster} and \textit{M. musculus}.
Figure~\ref{fig:5} shows the histograms of the family size distributions and the corresponding fits with the prediction of the neutral evolutionary model in  Equation~(\ref{eqx2}). 
 In all tested cases, the motif families distribution follows the predicted functional form with a level of  agreement comparable to the human case discussed above.  
 However, there is a clear trend of the fitted parameter $\theta$ to increase with complexity as measured by the number of TFs in the species (or alternatively by the total number of genes).  
 This trend is reported  in Figure~\ref{fig:5} and it is sublinear in the investigated window of TF repertoires.   
The definition of  $\theta\simeq \frac{1}{1+\mu/\lambda}$ indicates that this trend corresponds to a decrease rate of cis-innovation, with respect to the duplication rate,
as the complexity of the organism increases. 
\newline
\noindent
The value of $\theta$ intuitively represents the level of ``redundancy'', i.e., the tendency of TFs  to keep the same binding preferences. 
Actually, this parameter  can be used to quantify the retained redundancy of TF binding in a neutral evolution context:
the higher is the $\theta$ value,  the slower is the TF divergence with respect to the duplication rate.  
The limit  value $\theta=1$  implies that the distribution in Equation~(\ref{eqx}) becomes a power-law distribution  with  motif families. 
Figure~\ref{fig:5} (right-bottom) shows that this level of ``redundancy'' increases with the organism complexity as  measured with the total number of TFs. 
Note that we tested with extensive simulations that the value of $\theta$ is not in principle dependent on the 
total number of TFs (see Figure S2) if the rates are constant. 
This further confirms that Figure \ref{fig:5} captures a non trivial trend of the innovation dynamics with genome size. 
\section*{Discussion}
In this paper we addressed quantitatively the evolutionary dynamics of the transcription factor repertoire.
We introduced and discussed a classification that groups the TFs by reason of their binding preferences into what we call motif families.
Such an approach is sensitive to a fine divergence in regulation that would have been undetectable using the DBD taxonomy.
The evolution of the motif families proves to be well described by a simple neutral model that depends only on one free parameter $\theta$.
Ultimately $\theta$ accounts for the relevance of divergence between TFs with respect to retention of redundant copies.
It can be seen as a readout of the level of redundancy of TF binding preferences, which reports how much the regulatory system has been
shaped by duplication vs innovation.
\newline
\newline
\noindent
We devised two main deviations from the neutral scenario that seem to be due to opposite evolutionary pressures.
A positively selected over-expansion of some families that are associated to multicellularity evolution.
The inhibition of duplication for a specific set of ancient TFs, or ``singletons'', that can be traced back to their unicellular ancestors.
Looking at the motif family organization allows to tackle the evolution of transcriptional regulation and identify global trends in comparative genomics,
since it does not require to know in detail the transcription network, but just the TF binding preferences.
Moreover, the parameter $\theta$ of redundancy grants an easy way to compare different organisms.
\newline
\newline
\noindent
A major issue in the study of the evolution of regulatory systems is to identify those features which can be in some way associated to the organism complexity.
Combinatorial regulation is a distinctive feature of complex eukaryotes. 
Indeed, prokaryotic and eukaryotic TFs use different binding strategies, with PWMs of high and low information content respectively~\cite{wunderlich2009different}.
This difference is related to the evolution of the combinatorial strategies of control, typical of higher eukaryotes, that can compensate the low information content 
of their TF binding sites by combining several of them in the same promoter~\cite{wunderlich2009different}. 
This could have also  been favoured  by the widespread presence of transposable elements able to convey combinations of TF binding  sites all over the genome~\cite{Testori2012}.  
However, if the PWMs that characterize a motif family have low information content, 
the set of preferred binding sequences is loosely defined and can include several possible sequences. 
Thus, the mutation process is less likely to drive a TF away from its motif family.  
 This would translate in a lower  cis-innovation rate in our model for organisms with higher complexity, and this trend seems indeed to emerge from our comparison 
 of the different motif family organization in different species (Figure~\ref{fig:5}). 
\newline
\newline
\noindent
The increased  degeneracy of TF PWMs can also have another relevant consequence. 
Having larger motif families enables a different layer of combinatorial regulation, where several redundant TFs compete for the same binding site.
In other words, a binding site may be subject to the combinatorial regulation of several TFs as well as a promoter is subject to the combinatorial regulation of several binding sites.
Our findings suggest that eukaryotes of increasing complexity do not need only a richer repertoire of TFs to regulate an expanded genome, but also an increased
redundancy of TF PWMs. Speculatively, such an increase is aimed at the implementation of this additional layer of combinatorial regulation.
\newline
\newline
\noindent
In conclusion, complexity seems to be associate to the redundancy of the TF repertoire, i.e., to the presence of large families of TFs which recognize the same binding sequences. 
It would be interesting to understand the consequences of this observation on the topology and function of the regulatory network.
\section*{Methods}
\subsection*{Data set} 
We took advantage of the Catalog of Inferred Sequence Binding Preferences (CIS-BP database~\cite{weirauch2014determination},
\rev{version number 1.02} ), 
which collects the specificities of a vast amount of TFs in several species. 
The PWMs in this database were either directly derived from systematic protein binding microarray (PBM) experiments or inferred
by overall DBD amino acid identity.
Furthermore, the CIS-BP database gathers  data from all the main existing databases (such as TRANSFAC~(\cite{matys2006transfac}), JASPAR~(\cite{mathelier2013jaspar}) and SELEX~(\cite{jolma2010multiplexed})) and 
several Chip-Seq experiments, which had been used for cross-validation.
To construct the motif families, we downloaded the PWMs associated to each TF, considering both those obtained from experimental assays and the inferred ones.
In this way, we obtained 4172 PWM unique identifiers (PWD IDs) annotated to 906 different TFs.
\subsection*{The BDI model}
We define as ``class $i$'' the set of all families of size $i$.  
 $f_i$ represents the number of families in the i-th class and $M$ be the total number of classes $i=1....M$ corresponding to the possible family sizes, 
with  $M$ at most equal to the total number of elements $N$.  

The evolution equations are:

\begin{align}
 \frac{\mathrm d f_1(t)}{\mathrm d t}= & -(\lambda +\delta+\mu)f_1(t)+2(\delta+\mu) f_2(t) + \mu N + \nu \nonumber\\
 \frac{\mathrm d f_i(t)}{\mathrm d t}= & (i-1)\lambda f_{i-1}(t) -i(\lambda +\delta+\mu)f_i+ \nonumber\\
                                       & +(i+1) (\delta+\mu) f_{i+1}(t)\label{master_eq}\\
 \frac{\mathrm d f_{M}(t)}{\mathrm d t}= & (M-1)\lambda f_{M-1}(t) -M(\delta+\mu)f_{M}(t) \nonumber
\end{align}

where $\lambda$, $\delta$, $\nu$  and $\mu$ denote the birth, death, \textit{de novo} innovation 
and \textit{cis-}innovation rates respectively.

The model can be mapped in the simplest case of the BDI models discussed in ~\cite{karev2002birth} 
with the substitution $\delta^{\prime}=\delta+\mu$ and $\nu^{\prime}=\nu+\mu N$.

From the general solution discussed in~\cite{karev2002birth}, we obtain at steady state:

\begin{align}
 & f_i=\frac{\nu^{\prime}}{\lambda} \left(\frac{\lambda}{\delta^{\prime}}\right)^i \frac{1}{i} \sim \frac{\theta^{i}}{i}
\label{sol}
\end{align}

where $\theta=\frac{\lambda}{\delta^{\prime}}=\frac{\lambda}{\delta+\mu}$.
If, following~\cite{karev2002birth}, we assume a balance between birth and death rates $\lambda=\delta$ then 
$\theta=\frac{\lambda}{\lambda+\mu}$ and Eq.(\ref{sol}) becomes:

\begin{align}
 f_i=\frac{\nu+\mu N}{\lambda} \left(\frac{\lambda}{\lambda+\mu}\right)^i\frac{1}{i}
 \label{sol2}
\end{align}

The deviation of $\theta$ from 1 allows to estimate the magnitude of   $\mu$ with respect to $\lambda$. 
In the limit of $\theta \to 1$ ($\mu \to 0$) the usual power-like behaviour of the standard DBI model is recovered. 
Since we know $\nu\ll\mu$, we shall assume $\nu=0$ and the solution of the model eq.(\ref{sol2}) becomes a function only of $\theta$.
\begin{align*}
 f_i=N\frac{\mu}{\lambda} \left(\frac{\lambda}{\lambda+\mu}\right)^i\frac{1}{i}=N(1-\theta)  \frac{\theta^{i-1}}{i}
\end{align*}

\subsection*{Maximum Likelihood estimation of $\theta$}

To perform a MLE of the parameter $\theta$, we must first move from the distribution of the number of families to a probability distribution. 
This is simply achieved by normalizing the $f_i$.
$p_i=C_{M} \frac{\theta^i}{i}$. The normalization constant $C_M$ assumes a very simple form in the large $M$ limit:

$$C_M = [\sum_{i=1}^{M}{\frac{\theta^i}{i}}]^{-1} \stackrel{M \to \infty}{=} [-\ln(1-\theta)]^{-1},$$
leading to the probability distribution:
\begin{flalign}
 p_i= \frac{1}{-ln(1-\theta)}\frac{\theta^{i}}{i}
\label{prob}
\end{flalign}
We show in the Supplementary Material that for our range of values of $M$ and $\theta$ the error induced by this approximation is negligible.

The probability distribution in Eq. \ref{prob} is simple enough to allow an analytic determination of the MLE for $\theta$ (see the Supplementary Material for the detailed calculation), 
which turns out to be:

\begin{flalign}
 \theta_{MLE}=1-e^{\frac{1}{\overline{k}}+W_{-1}\left(-\frac{1}{\overline{k}}e^{-\frac{1}{\overline{k}}}\right)}
\end{flalign}

Where $\overline{k}$ is the mean size over the sample and W is the Lambert Function.

\subsection*{Goodness-of-fit test}

We compared the empirical data with our model,  defined by $\theta_{MLE}$, 
following the strategy proposed in~\cite{clauset2009power}. 
More precisely we used the Kolmogorov-Smirnov (KS) statistic as a measure
of the distance between the distribution of the empirical data and our model. In order to obtain an unbiased estimate for the p-value,
we created a set of one thousand 
synthetic data samples with the same size of the empirical one, drawn from a distribution with the same $\theta_{MLE}$ value.  
For each synthetic sample, we computed the KS statistic relative to the best-fit law for that set and constructed the distribution of  KS values.
The p-values reported in the paper represent the fraction of the synthetic distances larger than the empirical one.

\subsection*{Gene Ontology}

We performed a gene ontology analysis on the genes belonging to the union of the three larger motif families using the over-representation test of the PANTHER facility~(\cite{mi2013panther})
and selecting only the  Biological Process ontology. 
We chose as a background  for the test the entire data sample (906 TFs) to eliminate  annotations simply associated to  generic regulatory functions of TFs.  
p-values were evaluated using the Bonferroni correction.

\section*{Acknowledgements}

We thank F.D. Ciccarelli and M. Cosentino  Lagomarsino for critical reading of the manuscript, and 
 A. Colliva, M. Fumagalli and A. Mazzolini  for useful discussions.
\subsection*{Funding}The work was partially supported by the Compagnia San Paolo grant GeneRNet.

\section*{Author contributions statement}
 M.C conceived the project;   A.R., M.O. and M.C. analysed the data and developed the model. 
 \textcolor{red}{A.C. built and analysed the motif network.} All authors wrote and reviewed the manuscript. 

\section*{Additional information}

\textbf{Competing financial interests} The authors declare no competing financial interests. 


\bibliography{biblio}

\end{document}